# Colossal orbital Zeeman effect driven by tunable spin-Berry curvature in a kagome metal

Hong Li[1], Siyu Cheng[1], Ganesh Pokharel[2], Philipp Eck[3], Chiara Bigi[4], Federico Mazzola[5], Giorgio Sangiovanni[3], Stephen D. Wilson[2], Domenico Di Sante[6,7], Ziqiang Wang[1] and Ilija Zeljkovic[1]

[1] Department of Physics, Boston College, Chestnut Hill, MA 02467, USA.

[2] Materials Department, University of California Santa Barbara, Santa Barbara, California 93106, USA

[3] Institut für Theoretische Physik und Astrophysik and Würzburg-Dresden Cluster of Excellence ct.qmat, Universit ät Würzburg, 97074 Würzburg, Germany

[4] School of Physics and Astronomy, University of St Andrews, St Andrews KY16 9SS, United Kingdom

[5] Istituto Officina dei Materiali, Consiglio Nazionale delle Ricerche, Trieste I-34149, Italy

[6] Department of Physics and Astronomy, University of Bologna, 40127 Bologna, Italy

[7] Center for Computational Quantum Physics, Flatiron Institute, 162 5th Avenue, New York, NY 10010, USA

**Abstract:**

Berry phase and the related concept of Berry curvature can give rise to many unconventional phenomena in solids. In this work, we discover colossal orbital Zeeman effect of topological origin in a newly synthesized bilayer kagome metal $TbV_6Sn_6$. We use spectroscopic-imaging scanning tunneling microscopy to study the magnetic field induced renormalization of the electronic band structure. The nonmagnetic vanadium *d*-orbitals form Dirac crossings at the K point with a small mass gap and strong Berry curvature induced by the spin-orbit coupling. We reveal that the magnetic field leads to the splitting of gapped Dirac dispersion into two branches with giant momentum-dependent *g* factors, resulting in the substantial renormalization of the Dirac band. These measurements provide a direct observation of the magnetic field controlled orbital Zeeman coupling to the enormous orbital magnetic moments of up to 200 Bohr magnetons near the gapped Dirac points. Interestingly, the effect is increasingly non-linear, and becomes gradually suppressed at higher magnetic fields. Theoretical modeling further confirms the existence of orbital magnetic moments in $TbV_6Sn_6$ produced by the non-trivial spin-Berry curvature of the Bloch wave functions. Our work provides the first direct insight into the momentum-dependent nature of topological orbital moments and their tunability by magnetic field concomitant with the evolution of the spin-Berry curvature. Significantly large orbital magnetic moments driven by the Berry curvature can also be generated by other quantum numbers beyond spin, such as the valley in certain graphene-based structures, which may be unveiled using the same tools highlighted in our work.

**Main text:**

The Berry phase of electron wave functions can strongly influence the low-energy physics of solid state systems and give rise to various unconventional phenomena such as ferroelectricity, the topological Hall effect, and orbital magnetism [1–11]. Materials with flat bands and Dirac crossings are generally believed to be ideal places where such topological phenomena can be realized, in large part due to finite Berry curvature of the bands amplified at particular momentum-space positions. For example in Dirac systems, a small Dirac mass gap can be generated by the spin-orbit coupling, which in turn produces enormous Berry curvature concentrated at the momentum of the Dirac crossing. Lifting the degeneracy of the electronic bands characterized by large Berry curvature can lead to unusual magnetism originating from purely orbital contributions, which has so far been detected in certain exfoliated Moire structures [3]. Although Berry curvature effects permeate much of the physics of topological materials in condensed matter physics [1,2,4–11], spectroscopic insights into these phenomena have been challenging. In particular, while Berry curvature is a quantity intrinsically dependent on crystal momenta, experimental detection of ensuing phenomena with momentum resolution is yet to be achieved.

Kagome metals have emerged as a highly tunable platform that is ideal for the explorations of novel electronic phenomena intertwined with non-trivial band topology. The experimental efforts thus far have been focused on two different kagome metal sub-families. On the one side, magnetic kagome systems with Fe- [12–19], Mn- [20–24] or Co- [8,25,26] kagome nets, in which magnetic order of the transition metals in the kagome plane gives rise to various forms of magnetism, have been explored to study Dirac fermions [12,13], topological flat bands [8,27], Weyl fermions [26] and Fermi arcs [25,26]. These magnetic metals typically show a strongly spin-polarized electronic band structure, favorable for the realization of massive Dirac fermions and the Chern magnet physics [28] due to the strong time-reversal symmetry breaking associated with the magnetic order. Dirac fermion bands in these systems can be tuned by an external magnetic field, albeit the process is generally intertwined with spin canting effects such as that in $Fe_3Sn_2$ [13] and $YMn_6Sn_6$ [24]. On the other side, an entirely non-magnetic V-based $AV_3Sb_5$ (A=Cs, K, Rb) kagome metal family [29–31] with a spin-degenerate band structure attracted a tremendous attention [32–43] due to the emergence of novel density waves, exotic Fermi surface instabilities and unusual superconductivity.

The recently discovered family of V-based bilayer kagome metals in the $RV_6Sn_6$ ($R$ = rare earth ion) structure offers a new tunable physical regime between these two existing classes of layered kagome metals – magnetism can still be induced by the choice of the rare-earth element $R$, but the vanadium atom in the kagome layers remains non-magnetic [44–47]. Moreover, the $R$-induced magnetism does not seem to significantly affect the low-energy vanadium orbital derived band structure. Measurements have already demonstrated the robustness of the canonical features of the kagome systems including Dirac cones and flat bands [44,48,49], thus setting a fresh stage for the search for unconventional topological phenomena. In this work, we use spectroscopic-imaging scanning tunneling microscopy (SI-STM) in combination with first-principles electronic structure calculations to discover colossal orbital Zeeman effect of topological origin in $TbV_6Sn_6$. The observed orbital Zeeman splitting of the Dirac bands is fueled by the large spin-Berry curvature induced orbital magnetic moments coupling to external magnetic field. We find that the effect is surprisingly non-linear and that it changes concomitant with the evolution of the Berry curvature.

We focus on the bilayer kagome metal $TbV_6Sn_6$, which is a low-temperature soft ferromagnet ($T_c \approx 4.3$ K) due to ordering of the $Tb^{3+}$ magnetic moments [45,46]. The crystal structure of $TbV_6Sn_6$ consists of a V atom kagome net inserted between alternating $Sn^1$-$Sn^2$-$Sn^1$ slabs and $TbSn^3$ layers (Fig. 1a,b). We cleave bulk single crystals of $TbV_6Sn_6$ in ultra-high vacuum and immediately insert them into the STM head (Methods). STM topographs reveal a layered structure with $ab$-plane terraces on the surface, and step heights consistent with a single unit cell of $TbV_6Sn_6$ (Fig. 1c). We identify the termination predominantly observed in Fig. 1b,c as the $Sn^2$ termination due to the clearly resolved individual atoms forming a hexagonal lattice (top inset in Fig. 1d) and its similarity with the equivalent $Sn^2$ surface imaged in $YMn_6Sn_6$ [24].

Average d$I$/d$V$ spectra over a wide energy range show a spectral peak and a sharp upturn in conductance around 200 meV above the Fermi level (bottom inset in Fig. 1d). Temperature-dependent d$I$/d$V$ measurements reveal that spectra remain indistinguishable upon raising the temperature above the Curie temperature $T_c$ (Fig. 2a) and well below $T_c$ (Supplementary Figure 1), suggesting that the electronic structure in this energy range is not affected by the ferromagnetic ordering. Interestingly however, the spectra show a pronounced change as magnetic field is applied perpendicular to the sample surface (Fig. 2b,c). In particular, the spectral peak appears to split into two peaks that shift in opposite directions, i.e., one towards lower energy and the other conversely to higher energy (Fig. 2b,c). Magnetic field-induced spectral splitting is also robustly observed well inside the ferromagnetic state and well above $T_c$ (Fig. 2d), which again points to the evolution of electronic band structure largely unaffected by the localized moments of Tb atoms that saturate below 0.3-0.4 T [45,46] (Supplementary Note 2).

To gain further insight into the low-energy electronic band structure in momentum-space (*k*-space), we proceed to use spectroscopic-imaging scanning tunneling microscopy (SI-STM). The technique, also referred to as quasiparticle interference (QPI) imaging, relies on scattering and interference of electrons detectable as periodic conductance modulations in d$I$/d$V$(**r**,$V$) maps to provide an insight into the momentum-space electronic band structure. To obtain a high momentum-space resolution, performing measurements over large regions of the sample is desirable (Fig. 3a). We first focus on the spectroscopic mapping at zero magnetic field. d$I$/d$V$(**r**,$V$) maps show strong "ripples" in real-space centered at defect sites (Fig. 3b), reflected in an approximately isotropic wave vector $q_1$ seen around the center of associated Fourier transforms (FTs) (Fig. 3c). The wave vector magnitude changes with energy, which can be visualized in the radially-averaged dispersion shown in Fig. 3d. From this plot, we can observe that the imaged electronic band exhibits a nearly linear dispersion with velocity 1315 ± 64 meV·Å, starting at about 170 meV and extending over a large energy range to above 380 meV, where the signal disappears. We attribute the existence and the dispersion of $q_1$ to intra-band scattering arising from a gapped Dirac fermion dispersion at K (inset in Fig. 3d) for the following reasons. First, the dispersion velocity obtained from our data shows excellent agreement with the Dirac dispersion at K point appearing in a similar energy range in our theoretical calculations (Fig. 3d). Second, the morphology and the dispersion of $q_1$ is qualitatively similar to that measured in helical antiferromagnet YMn$_6$Sn$_6$ that has also been attributed to a Dirac dispersion at K [24]. Third, the QPI bandwidth of about 200 meV matches closely to the width of the upper part of the same Dirac dispersion in DFT calculations [50]. We note that the Dirac point imaged here at about +150 meV is different from that reported in Ref. [46], which focuses on the lower Dirac crossing positioned at -200 meV. Based on DFT calculations, the lower Dirac crossing is separated from the upper one by about 400 meV, which is nicely consistent with the two different Dirac dispersions imaged by ARPES [46] and STM (Fig. 3).

Next, we apply magnetic field parallel to the *c*-axis and measure the concomitant response of the Dirac dispersion at 4.2 K. Our field-dependent QPI measurements reveal that the Dirac band dispersion exhibits a pronounced field-induced change. Notably, the band appears to split into two branches – one shifting upwards to higher energy and the other one shifting to lower energy (Fig. 3d-h, Supplementary Figure 2, Supplementary Figure 3). This observation suggests that the Dirac band at zero magnetic field is spin-degenerate, consistent with the combined action of inversion and time-reversal symmetries. Similarly to the spectral peak saturation at magnetic fields $B \geq 4$ T in Fig. 2, the Dirac dispersion measured in the same range of magnetic fields appears indistinguishable (Fig. 3g,h), pointing to an intimate connection between the two observables. As the spectral peak emerges at approximately the same energy as the bottom of the massive Dirac band and shifts concomitantly with the band evolution (Fig. 4a,b), the spectral peak in Fig. 2 is likely a consequence of the large density of states near the Dirac band bottom, as also hypothesized to be the case in the Mn-based cousin, TbMn$_6$Sn$_6$ [28]. This interpretation is further confirmed by examining the field-induced shift of d$I$/d$V$ peaks and the QPI bands, which both exhibit a nearly identical behavior (Fig. 4e,f).

We stress that the field-induced band structure evolution observed here cannot be explained by spin canting, as reported for example, in kagome magnet $Fe_3Sn_2$ [13,18]. From magnetization measurements, it is clear that the magnetization in $TbV_6Sn_6$ reaches a plateau at a magnetic field of less than 0.3-0.4 T [45,46], so the saturation at much higher fields reported here is rather peculiar and cannot be explained in the framework of localized Tb magnetic moments that are already saturated at low fields. Further supporting the notion that the role of ferromagnetism from Tb moment ordering driving our observations is negligible, we point that the electronic band structure in the relevant energy range is indistinguishable in the ferromagnetic and paramagnetic states (Figure 2, Supplementary Figure 1, Supplementary Note 2).

This brings up an intriguing question of the origin of our observations. As the effect of local Tb magnetic moments can be largely excluded (Supplementary Note 2), we inspect itinerant orbital magnetic moments rooted in Berry curvature that can in some cases reach enormous levels [1,9,51]. We explicitly calculate orbital magnetic moments in this system associated with the massive Dirac dispersion imaged in our experiments. The large spin Berry curvature of the Dirac bands around the K point (see right inset of Fig. 1e for the spin up channel) prompts orbital moments of several meV/T (Fig. 4j, Supplementary Fig. 4). In addition, the orbital moment inversely depends upon the size of the Dirac masses, which can be theoretically adjusted by an artificial tuning of the local spin-orbit term. One can see that the maximum values of the orbital moment in the band region of interest change considerably, increasing as the Dirac masses decrease. Interestingly, the experimental band evolution is also not rigid – visual comparison near the band bottom reveals substantial rounding of the band curvature at higher fields (Fig. 4g). Our theoretical calculations suggest that both the gradual change in the effective $g$ factor with magnetic field and the anomalously large magnitude can be explained by the non-trivial Berry curvature effects associated with the Dirac band, giving rise to orbital magnetic moments. The non-linearity of the effect (Fig. 4e,f) may suggest that second-order corrections may be necessary to fully capture the underlying physics.

It is also worth noting that the magnitude of the band shift in our work is enormous (up to 11.20 meV/T), amounting to the magnetic moment on the order of 200 Bohr magnetons. This is more than two orders of magnitude higher than the conventional spin Zeeman effect $g$-factor $g_s \approx 2$ that would lead to 0.058 meV/T. It is also much larger than what has been detected from spectral peak shifts in $Co_3Sn_2S_2$ (0.078 meV/T [52] and 0.174 meV/T [8]), which were also attributed to orbital moments. Aside from the much larger magnitude reported here, our momentum-space evolution allows us to unambiguously determine the origin of these moments to arise from the Dirac band. At the quantitative level of comparison with our theoretical analysis, the experimental orbital moment is larger than the calculated one. Nonetheless, we stress here that the precise size of Dirac gaps and the strength of spin-orbit coupling might also be not well captured by density functional-based calculations, specifically when electronic correlations could be relevant [53,54].

Our experiments provide a direct visualization of orbital Zeeman splitting of topological origin and the first momentum-space insight into the surprisingly large orbital magnetic moments fueled by the enormous spin-Berry curvature at the gapped Dirac point. Our observations directly demonstrate that the opening of the Dirac gap in this system is not due to the time-reversal symmetry breaking magnetic order of the Tb atoms. It rather arises from strong spin-orbit coupling, which in turn leads to a non-zero spin Berry curvature. The latter generates the orbital magnetic moment $\pm \boldsymbol{m}(\boldsymbol{k})$, opposite for the two spin species but pointing in the *c*-axis. In a time-reversal symmetric situation, the opposite orbital magnetic moments for the two spin species would exactly cancel out; however, when time-reversal symmetry is broken by magnetic field, the two branches can become decoupled and split in external magnetic field. The applied magnetic field $\boldsymbol{B}$ along the *c*-axis couples to the dominating orbital magnetic moments by orbital Zeeman coupling $\pm \boldsymbol{m}(\boldsymbol{k})\boldsymbol{B}$ (ignoring the much smaller spin-Zeeman coupling) and lifts the degeneracy of the approximately spin-degenerate Dirac band. This allows the Dirac branches with opposite spins and opposite sign of the orbital moments to shift in opposite directions, as seen in our STM experiments.

We stress that our results here unveil a different physical mechanism compared to previous experiments imaging field-tunable electronic band structure [13,24]. In for example $Fe_3Sn_2$ [13] and $YMn_6Sn_6$ [24], the observed electronic band structure change inevitably happens in a regime where spins are continuously tilting with field, which makes it impossible to disentangle spin effects from orbital magnetic moment contributions, if any. $Fe_3Sn_2$ and $YMn_6Sn_6$ also exhibit a strongly spin-polarized electronic band structure with singly-degenerate Dirac dispersions composed of magnetic Fe or Mn *d* orbitals even in the absence of magnetic field. In our experiment on $TbV_6Sn_6$ here, we can conclude that the effects of Tb magnetic moments are minimal (Supplementary Note 2), which enables the momentum-resolved visualization of the elusive orbital physics and orbital Zeeman effect. Interestingly, the anomalously large magnitude of the orbital magnetic moments due to spin-Berry curvature in $TbV_6Sn_6$ may be attributed to the combination of a high Dirac band velocity and the relatively small spin-orbit coupling in a simple effective theory description (Supplementary Note 1). This could also provide a guiding principle to look for topological moments of even larger magnitudes in other materials, by focusing on maximizing the Dirac band velocity while minimizing the spin-orbit coupling induced Dirac gap. The striking evolution of effective magnetic moments from enormous to vanishing provides a surprising experimental insight into the phenomenological behavior of topological orbital magnetic moments strongly dependent on the Dirac band curvature and associated Dirac gap.

It is interesting to note that orbital magnetic moments that arise from Berry curvature of the Bloch electronic states can be generated by other quantum numbers. For instance, graphene-based systems are expected to also show orbital magnetic moments tunable by magnetic field. In single layer graphene, inversion symmetry breaking due to the attachment to the substrate creates a Dirac gap. However in contrast to $TbV_6Sn_6$, the relevant degree of freedom is the valley degree of freedom (K vs K'), not spin [55]. So in single layer graphene, there will also be Berry curvature induced orbital magnetic moments associated with gapped Dirac points, with the opposite sign at K and K'. External magnetic field could in principle couple to these, shifting one higher and the other one lower in energy, but there would be no orbital Zeeman splitting or spin Berry curvature. To visualize similar Berry curvature induced phenomena in graphene and Moire heterostructures, momentum-resolved magnetic field dependent imaging of the electronic band structure we presented in our work would be highly desirable.

**Methods**

**Bulk single crystal synthesis.** Single crystals of $TbV_6Sn_6$ were synthesized from Tb (pieces, 99.9%), Sn (shot, 99.99%), V (pieces, 99.7%) using a conventional flux-based growth technique. The flux mixtures of Tb, V, Sn were loaded inside a crucible with a molar ratio of 1:6:20 and then sealed in a quartz tube filled with Argon atmosphere. The sealed tube was heated at 1125 $^0$C for 15 hours and then slowly cooled down at a rate of 2 $^0$C/h to 780 $^0$C. Thin millimeter-sized shiny plate-like single crystals were separated from the molten flux via centrifuging at 780 $^0$C.

**STM experiments**. Single crystals of TbV$_6$Sn$_6$ were cleaved in ultrahigh vacuum and immediately inserted into the STM head, where they were held at ~4.5 K during measurements. STM data were acquired using a customized Unisoku USM1300 STM system. Spectroscopic measurements were made using a standard lock-in technique with 910-Hz frequency and bias excitation as detailed in the figure captions. STM tips used were home-made electrochemically etched tungsten tips, annealed in UHV to a bright orange color before STM experiments.

**Theoretical calculations.** We employed first-principles calculations based on the density functional theory as implemented in the Vienna ab-initio simulation package (VASP) [56], within the projector-augmented plane-wave (PAW) method [57]. The generalized gradient approximation as parametrized by the PBE-GGA functional for the exchange-correlation potential is used [58] by expanding the Kohn–Sham wave functions into plane waves up to an energy cutoff of 400 eV. We sampled the Brillouin zone on an 12x12x6 regular mesh by including SOC self-consistently. For the calculation of the spin Berry curvature and orbital magnetic moments, the Kohn–Sham wave functions were projected onto a Tb *d*, V *d* and Sn *s*, *p*-type basis [59], and the quantities were computed by using our in-house post-wan library [https://github.com/philipp-eck/post_wan], consistent with Ref. [60].

**Competing Interests**

The Authors declare no Competing Financial or Non-Financial Interests.

**Code availability**

The computer code used for data analysis is available upon request from the corresponding authors.

**Data Availability**

The data supporting the findings of this study are available upon request from the corresponding authors.

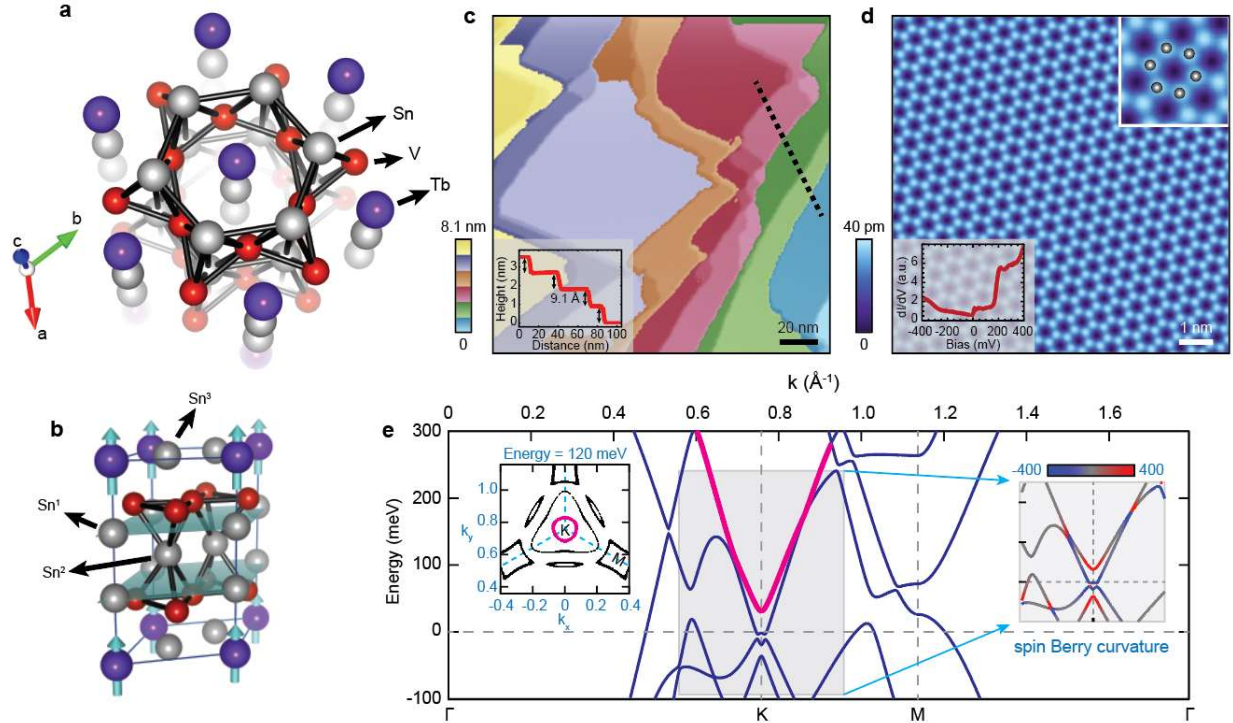

**Figure 1. TbV$_6$Sn$_6$ crystal structure and electronic band structure. (a)** Ball model representing the layer stacking order. **(b)** A 3D ball model showing the ferromagnetic structure of TbV$_6$Sn$_6$ (blue arrows represent the Tb spins pointing along the *c*-axis). **(c)** STM topograph of a 180 nm square region spanning multiple consecutive unit cell steps between Sn$^2$ hexagonal layers. The inset is the topographic line profile extracted along the black dashed line. Each step height is 9.1 Å, which equals the *c*-axis lattice constant. **(d)** STM topograph of Sn$^2$ termination. The upper right inset shows a zoom-in view of a small region and a hexagonal Sn lattice. The bottom left inset is a spatially averaged dI/dV spectrum taken in region in (c). **(e)** DFT calculations of TbV$_6$Sn$_6$ electronic band structure along the Γ-K-M-Γ line. The thick pink band is a gapped Dirac cone above the Fermi level, which carries a large spin Berry curvature for each spin channel (depicted in the right inset for spin up). More details of the relevant portion of the band structure with and without spin-orbit coupling can be found in Supplementary Figure 4. The left inset is a constant energy contour (CEC) around the K point at 120 meV. The pink circle is the CEC of the Dirac band at K. STM setup conditions: $I_{set}$ = 10 pA, $V_{sample}$ = 1V **(b)**; $I_{set}$ = 400 pA, $V_{sample}$ = 100 mV **(c)**.

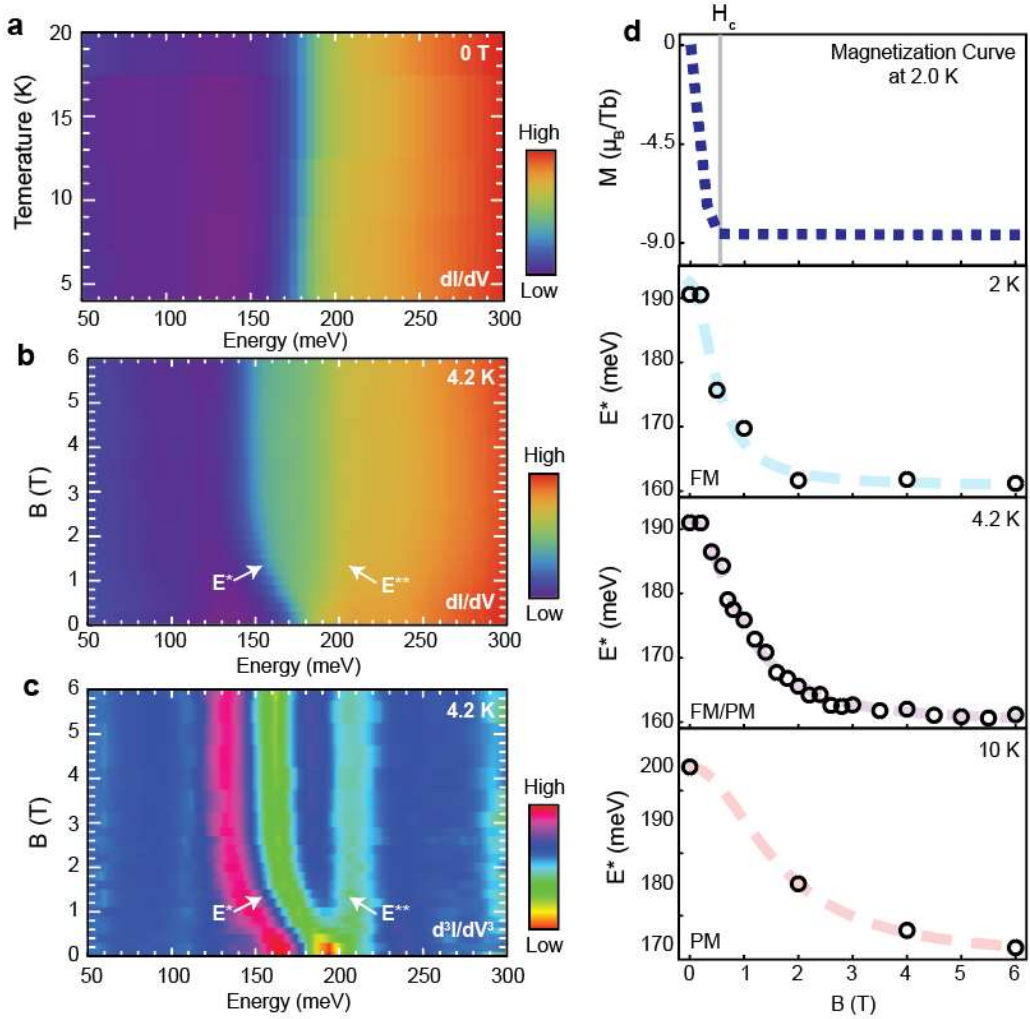

**Figure 2. The evolution of electronic density of states as a function of magnetic field and temperature.** (a) Spatially-averaged dI/dV spectrum as a function of temperature taken on the $Sn_2$ termination at 0 T. (b) Spatially averaged dI/dV spectrum and (c) associated second derivative of dI/dV as a function of magnetic field taken on the $Sn_2$ termination at 4.2 K. It can be seen that the dominant spectral feature at 0 T splits into two branches, at energies labeled as $E^*$ and $E^{**}$. Their dispersion saturates at higher fields. (d) Magnetization curve of $TbV_6Sn_6$ at 2 K (top panel) and evolution of $E^*$ as a function of magnetic field at 2 K (ferromagnetic, FM), at 4.2 K (near the transition point between FM and paramagnetic, PM phase) and at 10 K (PM) respectively. $E^*$ data points at 10 K were acquired over a different microscopic region of the sample tens of micrometers away compared to the 2 K and 4.2 K data; the 10 meV difference between $E^*$ at 10 K and those at lower temperatures is likely due to small chemical potential shift between different regions. Blue, purple and red dashed curves are visual guides denoting trends in the data. STM setup conditions: $I_{set}$ = 1 nA, $V_{sample}$ = 300 mV, $V_{exc}$ = 5 mV (a); $I_{set}$ = 1 nA, $V_{sample}$ = 300 mV, $V_{exc}$ = 2 mV (b,c).

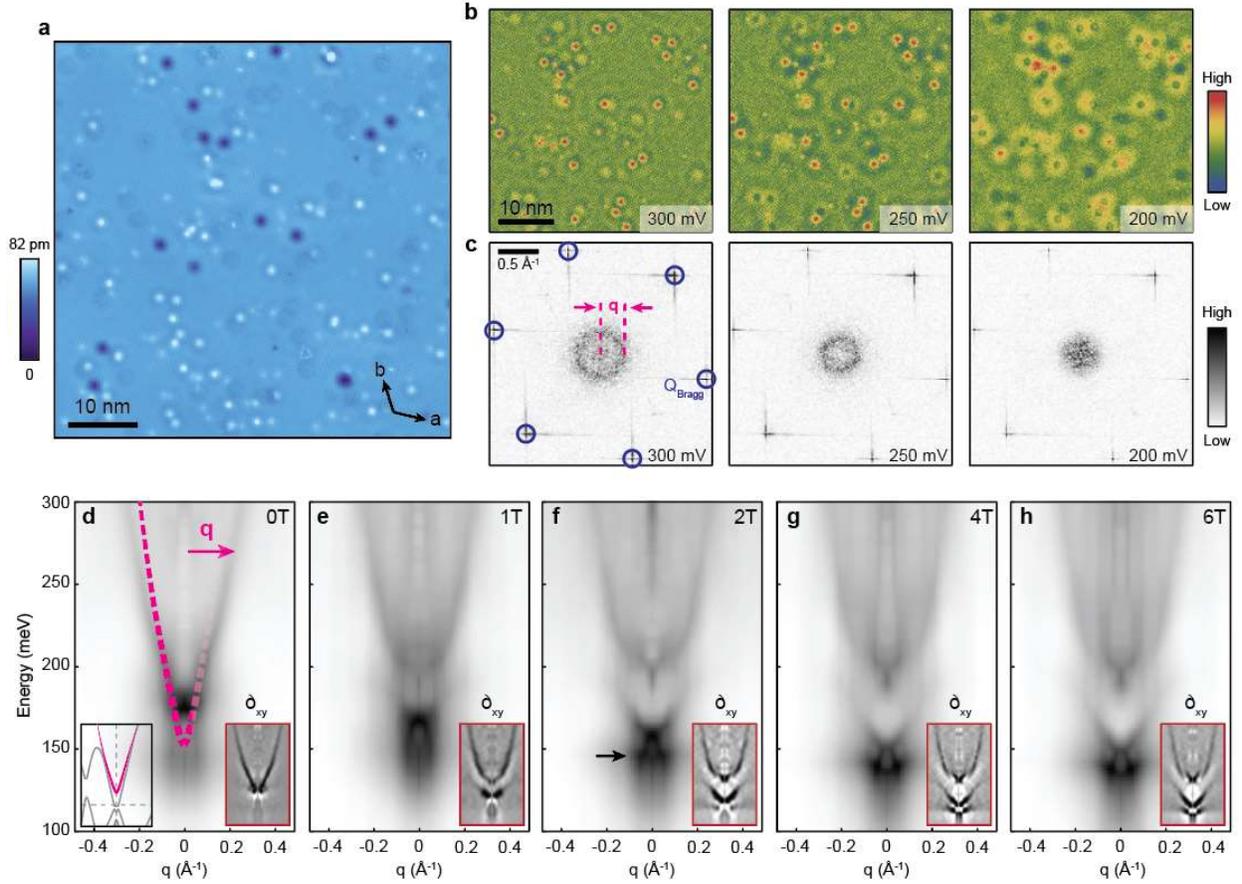

**Figure 3. Momentum-space visualization of the Dirac band using SI-STM. (a)** STM topograph of the Sn$^2$ hexagonal termination. **(b)** Representative d$I$/d$V$(**r**,$V$) maps of the region in (a) at biases 300 mV, 250 mV and 200 mV (from left to right panel) and **(c)** their Fourier transforms (FTs). Pink arrow denotes the dominant QPI wave vector **q** and the atomic Bragg peaks are circled in blue. **(d-h)** Energy-dependent radially-averaged linecuts $S(q,E)$, where $E$ denotes energy, starting from the center of the FTs in (c) at different magnetic fields applied perpendicular to the sample. The bottom right insets in (d-h) show the associated second derivatives of (d-h): $\partial_q \partial_E S(q,E)$. The bottom left inset in (d) refers to the DFT calculated Dirac band at K point slightly above Fermi level. By shifting the chemical potential down by 130 meV likely due to chemical doping by accidental defects, the Dirac band (dashed pink line in (d)) matches the STM QPI dispersion (gray background) very well. In addition to the Dirac bands, we also observe another wave vector at very low q that is nearly dispersionless but shifts with magnetic field (arrow in (f)), which we hypothesize it may originate from another band. Before $S(q,E)$ maps in (d-h) are generated, each d$I$/d$V$ map is processed by subtracting the average value to artificially suppress the bright center in the resulting FT and smoothed on 1 pixel length scale. We note that we apply the 1:2 conversion between momentum-transfer space ($q$-space) and momentum space ($k$-space) to superimpose the DFT-calculated band (pink dashed lined) on top of our STM data in (d). This conversion is applied because QPI wave vector **q** is a back-scattering vector, connecting **k** and –**k** points on the near-circular constant energy contour of the Dirac cone at K in momentum space. STM setup conditions: $I_{set}$ = 1 nA, $V_{sample}$ = 300 mV (**a**); $I_{set}$ = 1.5 nA, $V_{sample}$ = 300 mV, $V_{exc}$ = 2 mV (**b**). Data were taken at 4.2 K.

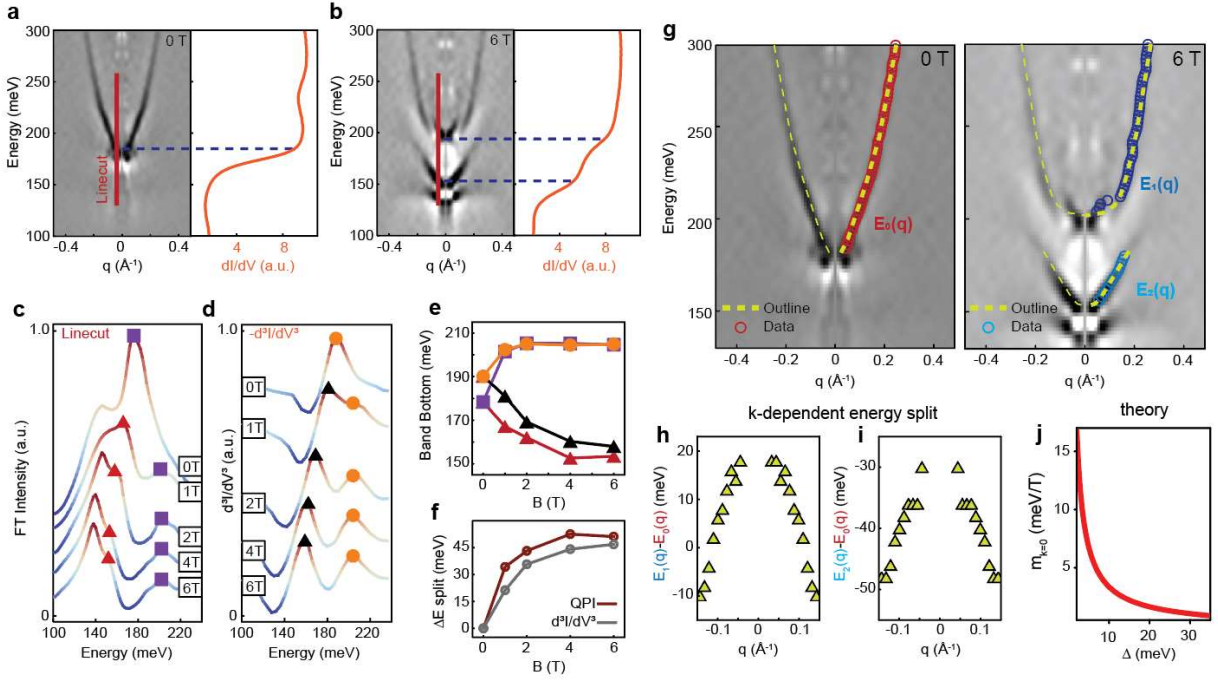

**Figure 4. Dirac band evolution and the saturation of the effective *g* factor. (a,b)** The second derivative of radially averaged linecuts in the Fourier transforms (FTs) of the d*I*/d*V* maps at 0 T (a) and 6 T (b). The right sides of the panels in (a, b) show the corresponding spatially averaged d*I*/d*V* spectra. The blue dashed lines denote the features in spectra likely corresponding to the band bottom. **(c)** The linecut taken along the red solid line in (a,b), as well as in equivalent datasets at 1 T, 2 T and 4 T. Purple squares and red triangles denote the band bottom positions in the energy axis. **(d)** Second derivative of d*I*/d*V* spectra from 0 T to 6 T. Orange circles and black triangles denote the peaks in the spectra. **(e)** Evolution of spectral markers with magnetic field extracted from (c,d) (color and symbol-coded to match the corresponding peaks in (c,d)). **(f)** Band splitting and spectral peak splitting extracted from (e). Brown line is extracted from QPI measurements representing band splitting, and the gray line is extracted from -d³*I*/d*V*³ curves representing the spectral peak splitting. **(g)** Radially averaged FT linecut starting at the center of the FT at 0 T and 6T (same as in insets in Fig. 3d-h), and the extracted band points denoted as $E_0(q)$ for the Dirac band at 0T, $E_1(q)$ for top Dirac band at 6 T and $E_2(q)$ for the bottom band at 6 T. The analysis of raw FT data leading to the same conclusion of Dirac band splitting can be found in Supplementary Figure 6. **(h,i)** The difference between energies of equivalent momentum-space states at different fields, obtained by the subtraction of $E_1(q)$ and $E_0(q)$ (h), and $E_2(q)$ and $E_0(q)$ (i). **(j)** Theoretically calculated evolution of the orbital moment *m* at the Dirac point (*k*=0) for the topmost Dirac cone upon a continuous change of the Dirac gap Δ. The gap is artificially tuned by acting on the strength of the local spin-orbit coupling in our first-principles calculations (Methods, Supplementary Figure 4). STM setup conditions: $I_{set}$ = 1.5 nA, $V_{sample}$ = 300 mV, $V_{exc}$ = 2 mV (a,b).

13